\documentclass[%
 reprint,
superscriptaddress,
showpacs,preprintnumbers,
 amsmath,amssymb,
 aps,
prl,
floatfix,
]{revtex4-1}

\usepackage{graphicx}
\usepackage{dcolumn}
\usepackage{bm}

\begin{document}


\title{Interdependence Theory of Tissue Failure: Bulk and Boundary Effects}

\author{Daniel Suma}
\affiliation{Department of Bioengineering, University of Minnesota}
\author{Aylin Acun}
\affiliation{
Bioengineering Graduate Program, University of Notre Dame}
\author{Pinar Zorlutuna}
\affiliation{
Department of Aerospace and Mechanical Engineering, University of Notre Dame}

\author{Dervis Can Vural}
\email{dvural@nd.edu\\}
\affiliation{
 Department of Physics, University of Notre Dame, USA}

\date{\today}

\begin{abstract}
The mortality rate of many complex multicellular organisms increase with age, which suggests that net aging damage is accumulative, despite remodeling processes. But how exactly do little mishaps in the cellular level accumulate and spread to become a systemic catastrophe? To address this question we present experiments with synthetic tissues, an analytical model consistent with experiments, and a number of implications that follow the analytical model. Our theoretical framework describes how shape, curvature and density influences the propagation of failure in a tissue subject to oxidative damage. We propose that aging is an emergent property governed by interaction between cells, and that intercellular processes play a role that are at least as important as intracellular ones.\end{abstract}

\maketitle

\section{Introduction}

As an organism ages, its cells shrink or enlarge, increase their lipid and pigment content, and lose their functionality and ability to proliferate. Mechanical theories of aging typically focus on the biomolecular mechanisms governing metabolism, cell damage and repair, such as oxidative stress, shortening telomeres, and various genetic factors \cite{murabito,rattan,shay,weinert,blackburn}.

However, aging is a complex phenomenon that spans multiple time and length scales. Organisms do not ultimately die because they run out of cells, but because cellular damage manifests as larger scale problems in tissues and organs, through a cascade of interactions. It remains unclear how failures dynamically propagate and accumulate to lead to frailty, aging related diseases, and ultimately, death \cite{dcv,boonekamp, nick,aylin,rockwood}.

The catastrophic end state, universality of survival statistics, and the irreversible nature of failure can be understood in terms of the dynamics of a network of interdependent components \cite{dcv}. According to this interdependence network theory of aging, if a component fails, then any other component that crucially depends on it will also fail. As a result, in well-connected interdependence networks, small failures will cascade into larger ones, and the probability of a system-wide catastrophe will monotonically increase with time, consistent with experimentally observed survival curves. The theory also predicts that the catastrophe is unavoidable even when repair rates far exceed the damage rate.

Some experimental work has been done, especially in cancerous tissues, to determine the effects of inter-cellular communication and stress \cite{lavi,kubo,greene,mehta,wang}. These cooperative effects, our earlier experiments \cite{aylin} and novel experiments reported here motivate us to view a tissue as a simple interdependence network where cells influence their neighbors' response to damage via various diffusive factors. 

Our aim here is to put forth a quantitative theory of tissue failure that is consistent with experimental data. In order to bridge microscopic (cellular) damage and macroscopic (organismic) catastrophes, we study the failure dynamics of an intermediate structure, the tissue. Specifically, we determine the relationship between damage propagation in functioning healthy tissues and the global and local properties, such as shape and density.

While the postulates of the quantitative model presented here are motivated by experiments on mammalian cells, we might expect similar results to also hold for eukaryotic colonies and bacterial biofilms where the survival of cells are linked to one other through a number of signaling and cooperative factors \cite{stoodley}. Thus, our results can be interpreted more generally, as the spatial dynamics of a cooperative population.

This paper is organized as follows: we first present experimental data on synthetic tissues that form the basis of our quantitative model. We then present analytical and computational results describing how failure propagates across a tissue.

\section{Results}
\subsection{Motivating Experiments}
Rat fibroblasts were encapsulated within hydrogels of differing geometries, in particular, a triangular prism with 30, 60, 90 degree corners and also a ``flat'' cylinder. The distance between cells were measured to be quite uniformly distributed around $\sim40\mu m$, much larger than the size of a fibroblast $\sim10 \mu m$.

The cell-laden hydrogels were subjected to continuous oxidative stress through cell culture media containing hydrogen peroxide (0.2 mM). Throughout the duration of the culture, live and dead cells were stained and counted in order to measure the death rate at the edges, corners, and bulk of the cell-laden hydrogels. The hydrogel material, which is Arginine-Glycine-Aspartic acid (RGD) conjugated polyethylene glycole 4-arm acrylate (PEG-RGD), was chosen to prohibit cells from migrating or proliferating \cite{yue, acun}. This way we ensured that the observed effects are not due to a loss or gain of physical contact between cells, but due to intercellular cooperation via factors secreted by the cells that diffuse through the hydrogel medium. Furthermore, when the cell density is sufficiently low, the aging effect disappears since the intercellular distance prohibits the interactions \cite{aylin}. In this dilute limit the death rate increases, further supporting our hypothesis that aging is driven by intercellular interactions rather than limitations in the externally provided resources.

We observed that cells located at smaller angles die faster than those at larger angles (Fig 1,2). We also observed that the cells encapsulated in a triangular geometry died faster than that encapsulated in a disk geometry (Fig1, top). In addition, cells started dying close to the edges and corners, and the dead layer propagated inwards. Specifically, when we plot the dead layer thickness a function of time for the 30, 60, 90 and 180 degree cases, it was seen that the smaller angles experienced faster propagation than the larger ones (Fig 1, bottom). 

These observations are unlikely to be caused by a diffusion limitation since the nutrient and oxygen diffusion penetration length is much larger than the height of the tissue, and thus, oxygen and nutrients are in excess. Furthermore, the geometry of the samples are effectively two dimensional planes, allowing every cells to receive equal amounts of external resources, which mostly diffuse from the top and bottom inward. Note that if the transport of nutrients and waste were the primary problem, we would expect to see the center die first and propagate outwards, the opposite of what we observed. 

The second, and perhaps more interesting reason that allows us to hypothesize that there is a cooperative interaction between cells is the following: If cells died \emph{only} due to oxidation, then we would see a constant rate of death (e.g. $5\%$ of the remaining live cells would die every day). However we observe an accelerated rate of death (e.g. every day a larger percentage of the remaining live cells die). This suggests that lack of cells contribute an \emph{extra} death rate in addition to oxidation and other single-cell-level causes of damage. Our earlier experiments do not exhibit a time dependent death rate, when cells were separated enough that they could not interact \cite{aylin}.

It should be noted that the increased fluorescence around the edges of the hydrogels is due to swelling and induced edge curvature from the manufacturing process.  The fluorescent level was not used to count cells as only individual points were counted and remapped to density based heat maps. 

\begin{figure}
\includegraphics[width=\linewidth]{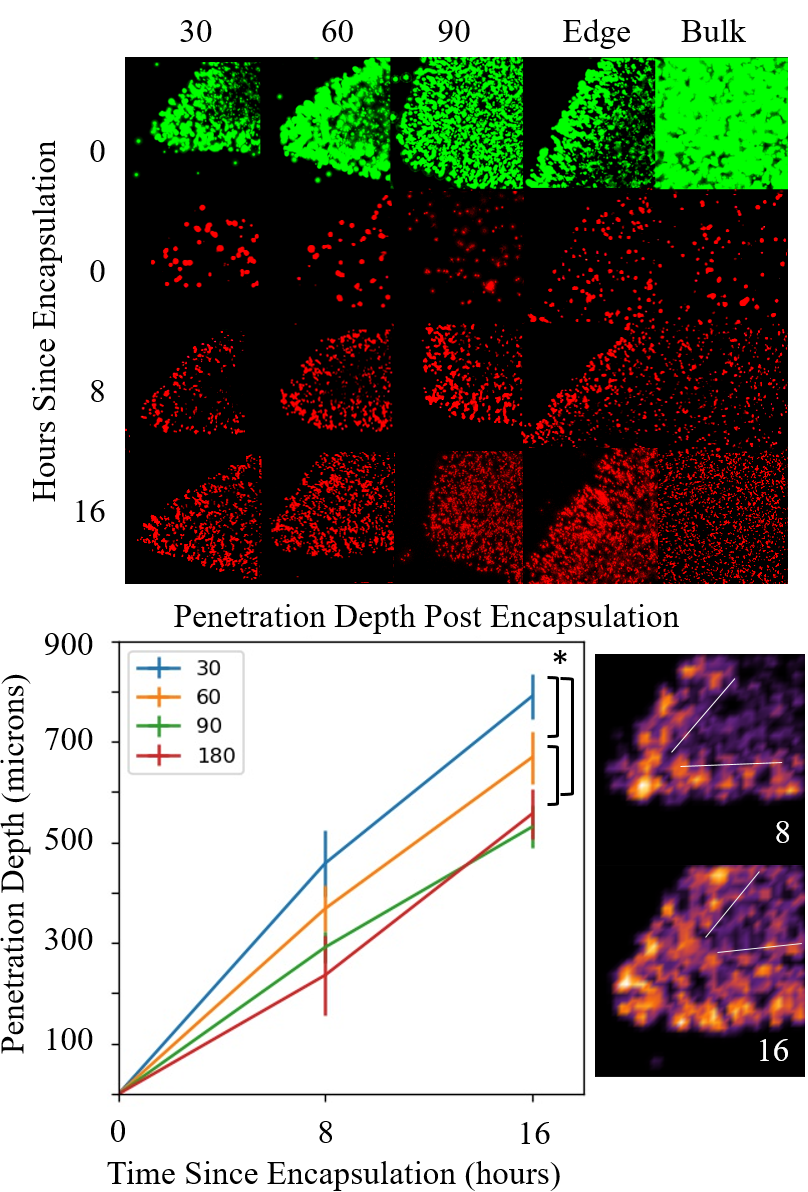}
\caption{{\bf Dependence of Population dynamics on Geometry}.  Top panel: Images taken at time points 0, 8, and 16 hours (top to bottom) after the start of stress treatment which show the collapse of each region  of the geometry. Live cells (shown in green, first column) marks the initial condition and dead cells (shown in red, remaining columns) marks dead cells at the specified time points. A layer of dead cells form and rapidly thicken until tips are completely consumed and the population collapse propagates inwards. Plotted in the bottom left panel is the penetration depth, i.e. the thickness of the dead cell layer, as a function of time for the 30, 60, 90 and 180 degree cases. We obtained the penetration depth from quantifying the dead cell number density at time 8 and 16 hours (bottom right panel), delineated the region separating the dead cell layer from the live cells (shown as thin white lines), and measuring the thickness of this layer from the tip. We used the dead cell number density as opposed to light intensity because edges appear brighter in the microscope images due to swelling of the edges, and not an increased dead cell population. The statistically significant differences has been marked with *, where a two tailed T test with unequal variance gives $p$ values of $1.6\times10^{-4}, 0.012$, and $0.006$ for red vs. blue, red vs. green, green vs. blue curves respectively, at the 16 hour mark.
}
\end{figure}

\subsection{Analytical Theory}
To make sense of the geometry dependence of tissue lifetime and population dynamics we propose an analytic model, the assumptions of which, stated qualitatively, are as follows: {\bf(1)} There exist damaging agents in the inter-cellular environment, and the cells counter this by secreting diffusive factors. Here we refer to the latter as ``cooperative factors'' (CF). The effect of CF's may be both direct or indirect, for example, neutralizing oxygen species by directly binding to them, catalyzing or promoting intermediate byproducts that react with them, or participating in reactions or reaction cascades that activate higher order cell repair mechanisms. {\bf(2)} The probability of death of a cell is proportional to the relative abundance of the non-neutralized damaging agent, as given by first order reaction kinetics. {\bf(3)} Cell lifetimes are much larger than the time it takes for the CF to diffuse throughout the sample, and the decay time of the molecule. 

To be more precise, we assume that the concentration $\phi_i$ of CF at position $r$ secreted by a single cell $i$ located at $r_i$, is approximately governed by the diffusion equation,
\begin{eqnarray}\label{dif}
\frac{\partial \phi_i(r)}{\partial t}=D\nabla^2\phi_i(t)-\gamma\phi_i(r)+A \,\delta(r-r_i)
\end{eqnarray}
where $\gamma$ and $D$ are the decay rate and diffusion constant of the CF in the extracellular matrix. The total concentration of CF experienced by cell $i$ is  then given by the sum of secretions by all cells $\Phi(r_i)=\sum_j\phi(|r_i-r_j|)$.

Secondly, given a local concentration $\Phi(r_i)$ near cell $i$, its probability of death per unit time is assumed to have a Michaelis-Menten-Hill form,
\begin{eqnarray}\label{hill}
P_i=\frac{\alpha\Phi_0^k}{\Phi(r_i)^k+\Phi_0^k}
\end{eqnarray}
where $\Phi_0$ is a constant that characterizes a threshold of cooperative factor concentration below which the cell's survival is compromised, and $k\geq1$ is the Hill constant, determined by the stoichiometry of the reactions in which $\phi$ is consumed. $\alpha$ is a proportionality constant that connects the probability of death to the concentration of cooperative factors or some unspecified molecule or structure that reacts with it.

We will now analytically solve the survival characteristics of the tissue in the bulk under certain reasonable approximations, and also simulate the system. The steady state--solution of (\ref{dif}), with appropriate boundary conditions, gives the influence of all surrounding cells on cell $i$
\begin{eqnarray}\label{totalcf}
\Phi(r_i,t)=\frac{A}{4\pi D}\int_\Omega\frac{e^{-\lambda|\vec{r}-\vec{r}_i|}n(\vec{r},t)}{|\vec{r}-\vec{r}_i|} d^3r.
\end{eqnarray}
where $n(\vec{r},t)$ is the cell density at a given position and time. Every cell experiences a different amount of CF depending on the number of functional cells that surround it. In general, those near the boundary of the tissue have lesser neighbors. To calculate the overall loss of population, we calculate the cooperative factors received, averaged over position,

\begin{align}
\langle\phi(r_i,t)\rangle_i=\frac{1}{N}\int_\Omega n(r,t)\phi(r,t)d^3r
\end{align}
where $N$ is the total number of cells. In the limit  $\lambda L\ll1$, this expression is analogous to the problem of finding the electrostatic energy of a spherical charge distribution. In this limit, the result is well known, $\langle\phi(r_i)\rangle_i=\frac{A}{4\pi}6N/(5L)$. For an arbitrary $\lambda L\equiv d$ we obtain
\begin{align}\label{ave}
&\langle\phi(r_i,t)\rangle_i=\beta\Phi_0 N(t)/N_0\\
\beta&\equiv 6AN_0[6-e^{-d}(6+6d+3d^2+d^3+d^4/4)]/(\pi Ld^5D\Phi_0)\nonumber
\end{align}
for a uniform cell density $N\equiv 4\pi L^3n(t)/3$. As the cells die near the boundaries with higher likelihood, the uniformity assumption will break down. However the agreement between discrete simulations and analytical theory indicate that the assumption of uniformity introduces a small overall error. We expect this error to be proportional to the surface area to volume ratio of the tissue.

The population of the cells is determined by their rate of death, $dN/dt=-PN$, which can be rewritten by substituting the approximation (\ref{ave}) into (\ref{hill}),
\begin{eqnarray}
\frac{dN(t)}{dt}=\frac{-\alpha N(t)}{1+(\beta N(t)/N_0)^k}.
\end{eqnarray}
This equation can be solved exactly:
\begin{eqnarray}\label{n}
N(t)=\frac{N_0}{\beta}\left[W(\beta^ke^{-\alpha k t+\beta^kN_0^k})\right]^{1/k}
\end{eqnarray}
where $N_0$ is the initial population of the tissue, and $W(x)$ is the Lambert function, recursively defined by $W(x)=\log(x)-\log(W(x))$ and can be expanded, $W(x)\sim\log(x)-\log(\log(x)-\log W(x))\ldots$. 

In Fig.2, we compare this theoretical result with experimental young and ``old'' cell populations \cite{aylin}. The former group was obtained by acclimating neonatal rat cells in oxidative media for two days, whereas the latter group was obtained by repeated cell division of the neonatal cells till they reach the Hayflick limit. The population of both groups, encapsulated in a 3D hyrogel, were measured every day. In both cases, we see that the qualitative shape of the survival curves are governed by the interdependence effect, although the old cells die quicker. The agreement between theory and experiment is acceptable.

In Fig.3 we directly simulate the population dynamics of a tissue consisting of cells with random positions to compare it with our approximate analytic formula (\ref{n}) and find good agreement.
\begin{figure*}
\includegraphics[width=\linewidth]{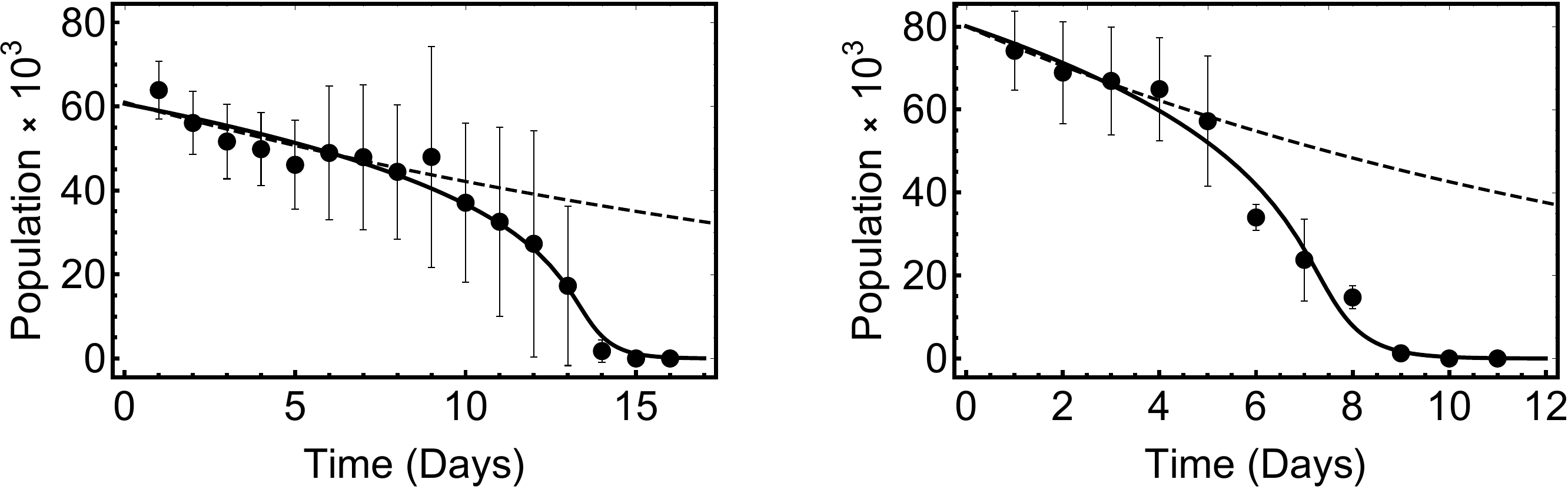}
\caption{{\bf Experiment (data points) versus theory (solid curve) versus null model (dashed line)}. The survival of cells acclimated in hydrogen peroxide for two days (left) and cells brought to Hayflick limit (right). Parameters used for the equation are $\beta=3.8, \alpha=1.5, k=3$ (left) and $\beta=3.1, \alpha=1.5, k=3$ (right). For comparison purposes we have also included the behavior of the null model where cells die independently of each other at a constant rate (dashed curve). Since the physiological requirements and biochemistry in both experiments should be similar, we have fixed $k$ and $\alpha$ and obtained the fits by varying $\beta$ using mathematica's in-built chi-square test. The $p$ values for the fits are $5\times10^{-15}$ and $1\times10^{-24}$. The moderate difference in the $\beta$ value could be due to a difference between young and old cells in thir cooperative factor secretion rates $A$, or their ability to receive / process cooperative factors $\Phi_0$ }
\end{figure*}

To obtain the bulk lifespan $\tau_b$ of the tissue,  we will first approximate the Lambert function by only keeping the leading term in its recursive definition, $W(x)\sim\log(x)$ (the accuracy of this approximation is demonstrated in Fig.4 with red dashed lines). This gives us,
\begin{eqnarray}\label{approx}
N(t)\sim N_0\left(B-C t\right)^{1/k}
\end{eqnarray}
where $B=1 + k\log(\beta)/\beta^k$ and $C=\alpha k/\beta^k$ (note that when $\beta$ is large $B\sim1$ and $N(0)\sim N_0$, as it should be). Then, from the approximate form (\ref{approx}), the lifetime $\tau_b$ can be estimated by setting $n(\tau_b)=0$,
\begin{eqnarray}\label{tau}
\tau_b=\frac{\beta^k+k\log\beta}{\alpha k}
\end{eqnarray}
Under the same approximation, the cell mortality rate (probability of death per unit time) is
\begin{eqnarray}
-\frac{1}{N(t)}\frac{dN(t)}{dt}\equiv\mu(t)=\frac{\alpha}{1+\beta^k+k \log\beta-\alpha k t}
\end{eqnarray}
These approximate and exact analytic results are plotted in Fig.3 and Fig.4. The former shows the agreement between analytical theory and simulations.

\begin{figure*}
\includegraphics[width=\linewidth]{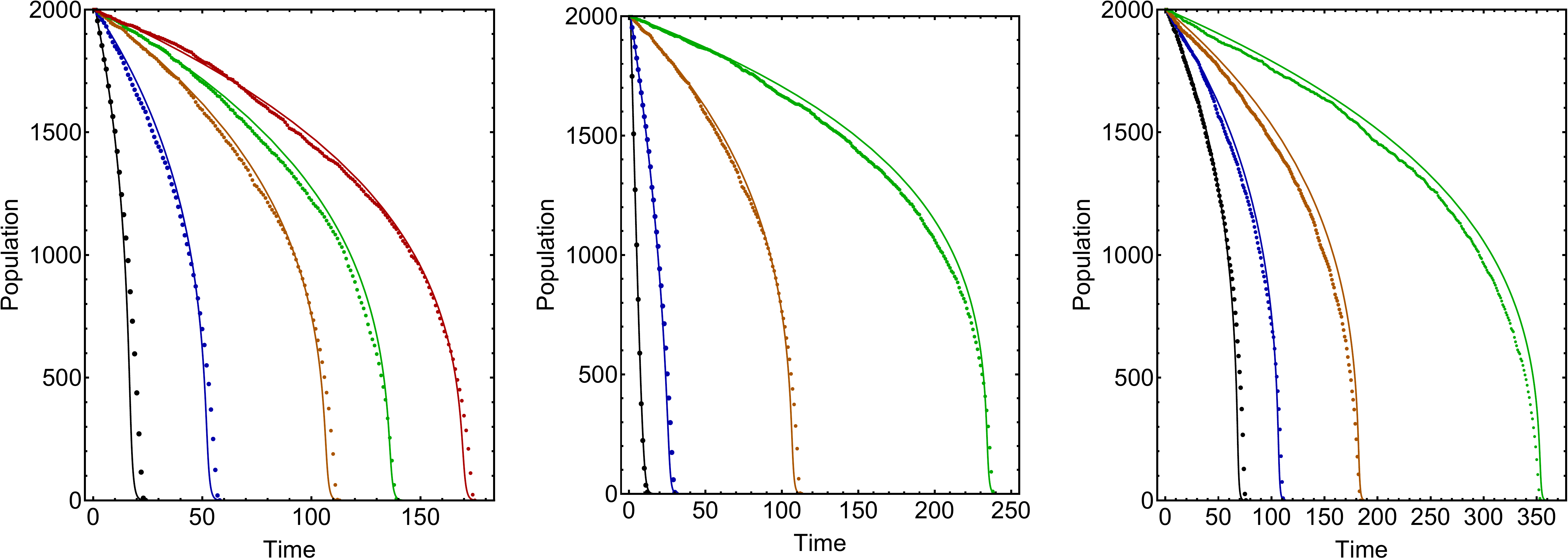}
\caption{{\bf Dependence of Population Curves on Diffusion Length $1/\lambda$ (left), Hill Coefficient $k$ (center), and required cooperative factor threshold $\Phi_0$ (right).} We compare simulations (dots) with analytical theory (solid line) for $1/\lambda$= 1,2,5,10,100 (black, blue, orange, green,red) units (left panel); and $k=1,2,3,3.5$ (black, blue, orange, green) (center panel); and $\Phi_0=350, 300, 250, 200$ (green, orange, blue, black) (right panel). Remaining parameters are kept constant at $N=2000$, $L=1$, $\Phi_0=300$, $k=3$ and $1/\lambda=5$ units.}
\end{figure*}

Since the cells located on the surface of the tissue have less neighbors, and thus less cooperative factors, we should expect failure to propagate from the surface towards the bulk, as observed in the experiments we report. If the radius of curvature of the shape of the tissue is much larger than the diffusion length $1/\lambda$, a cell at the surface will be exposed to half of the cooperative factors relative to a cell in the bulk. Thus the population density near the surface $n_s(t)$ can be found by simply letting $\beta\to\beta/2$ in identities (\ref{n}) and (\ref{tau}).
\begin{figure*}
\includegraphics[width=\linewidth]{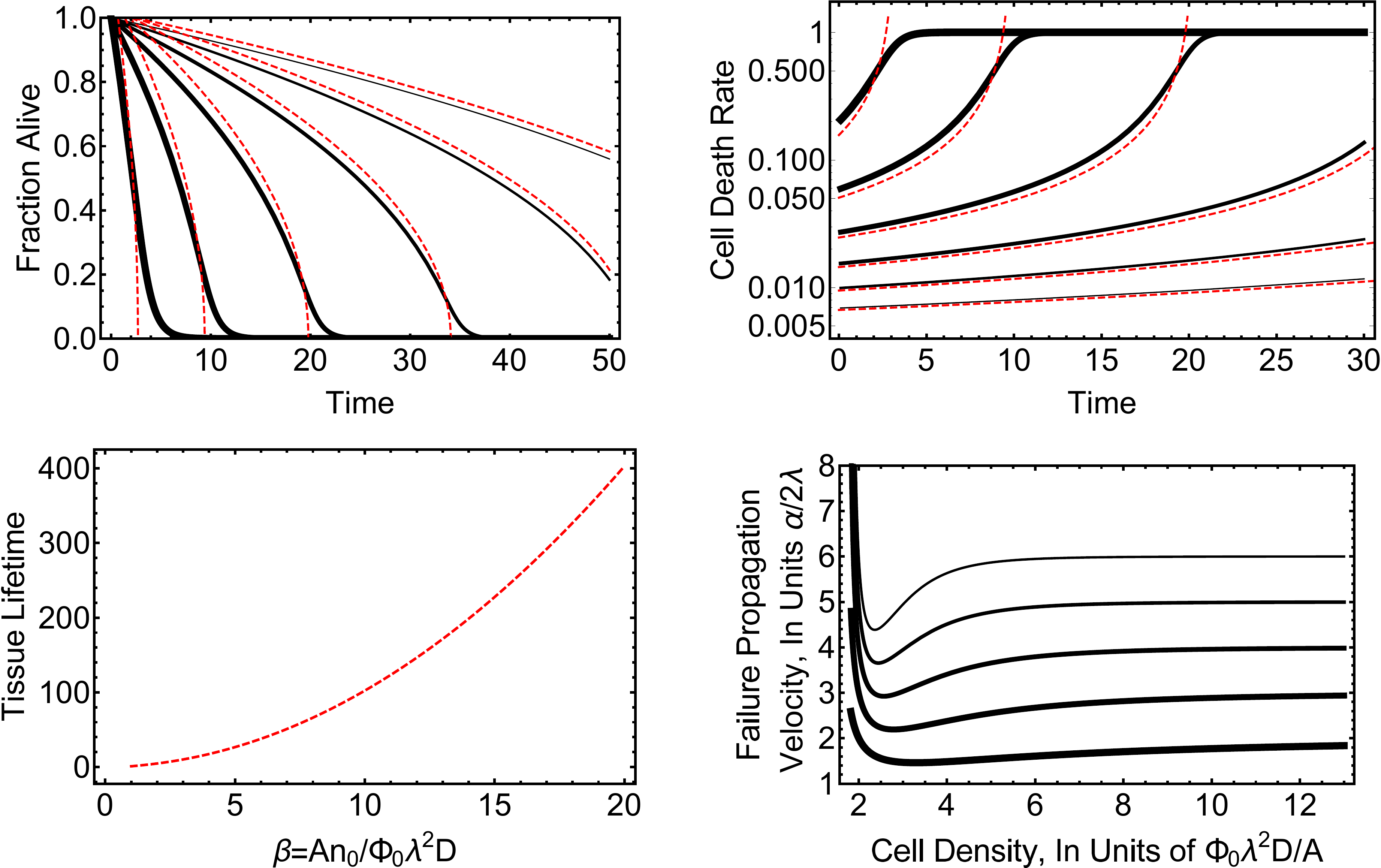}
\caption{{\bf Population Dynamics of a Strongly Interacting Tissue Model}. In all panels the solid black curves are the exact analytic solutions, whereas the dashed curves are approximate solutions based on eqn(\ref{approx}). In the top left and top right panels we plot the fraction of surviving cells and cell death rate  as a function of time. The system parameters are $\{\alpha,k,N_0\}=\{1, 2, 1000\}$, and the solid curves, from thick to thin correspond to $\beta=2,4,6,8,10$. In the bottom left and bottom right panels we display the approximate formula for tissue lifetime as a function of $\beta\approx AN_0/(\Phi_0\lambda^2D)$, where $A$ is the CF secretion rate, $N_0$ the initial population, $\Phi_0$ the threshold CF level, $1/\lambda$ the CF decay length, and $D$ the diffusion constant of the CF in the medium. In the bottom right panel we plot failure propagation velocity from the surface to bulk as a function of cell density. 
}
\end{figure*}

We will now estimate the velocity $v=dx(t)/dt$ of failure propagation from the surface of a tissue of thickness (or radius) $L$ when $L\gg1/\lambda$. At $t=0$ the cells close to the surface have a lifetime of $\tau_s=\frac{1}{\alpha k} + \frac{\log(\beta/2)}{\alpha (\beta/2)^k}$. Once this initial layer dies, we have a new boundary and the same process repeats itself with a new and lesser $N_0=N(\tau_s)$. This process can be iterated in continuous time, up until time $t$ at which the bulk collapses, or the implosion is complete. i.e. whichever is met first: $t=\tau_b$ or $x(t)=L$. We obtain the death propagation velocity as $v\equiv dx/dt\approx (2\lambda)^{-1}/\tau_s(n)$. That is,
\begin{eqnarray}\label{velocity}
v(t)=\frac{\alpha}{2\lambda}\left(\frac{1}{k} + \frac{\log(b(t)/2)}{(b(t)/2)^k}\right)^{-1}.
\end{eqnarray}
where $b(t)=An(t)/(\Phi_0\lambda^2D)\sim AN_0(B-Ct)^{1/k}/(\Phi_0\lambda^2D)$ is defined analogously with $\beta$. 

Eqn(\ref{velocity}) can be integrated numerically to yield the failure depth $x(t)=\int_0^t v(t)dt$ as a function of time. Unsurprisingly the velocity is  nonzero at all times; however inspecting (\ref{velocity}) reveals a non-trivial density dependence: At high cell densities the velocity is constant, independent of $n$. As $n$ decreases $v$ decreases, and reaches a minimum value. As $n$ continues to decrease, $v$ increases again, and diverges at some critical density that makes the paranthesis zero (Fig 4). This is the critical density at which bulk death dominates surface death.

Organs have different shapes, and parts of a given organ can have varying degrees of curvature. In order to determine how the shape and curvature of a tissue affects its lifetime, we studied three idealized geometries. A tetrahedron, a cube and a sphere were formed with the same initial cell density and total cell number. The population dynamics of these three geometries were then simulated. The survival curves for these geometries can be seen in Fig.5.

Since sharp corners are more likely to decay due to the lack of neighbors, the ``live surface'' becomes progressively rounder as the failure propagates inwards. In the mean time, cells in the bulk also thin down, and depending on the simulation parameters, may or may not reach the critical density before boundaries collapse inward.

\begin{figure}
\includegraphics[width=\linewidth]{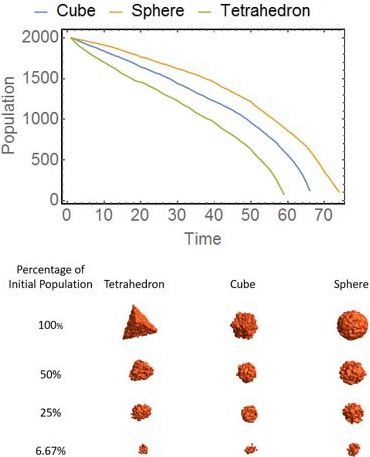}
\caption{{\bf Population Dynamics of Different Tissue Geometries}.  All shapes had identical volume and initial cell density. We used simulation parameters $k = 5$, $\Phi_0 = 175$, and $\lambda=2$.
}

\end{figure}

\section{Discussion}

Aging is often attributed to microscopic mechanisms that cause failure in the cellular level.  Our analytical, computational and experimental results support the view that aging is due to the failure of intercellular processes as much as it is due to intracellular ones. 

We have studied the details of how interactions between cells lead to the failure of the tissue. We have observed, through experiments and simulations, that damage starts from the boundaries of the tissue, and propagates inwards. At edges, external planes, and points, cells have lesser neighbors compared to the bulk, and thus receive less cooperative factors. Since the number of factors received by a cell is proportional to its probability of survival, failure originates near the boundaries and propagates inwards. In addition, we found that the inward collapse is the fastest for geometries with largest curvature or corners with the smallest angle. This trend was demonstrated experimentally and theoretically.

We have also seen that the larger death rate at the corners and edges causes the live cell surface to round up. The rounder boundary, in turn, decelerates the curvature, and thus, the local death rate. Experimentally a qualitative curvature/angle dependence of the collapse rate was demonstrated for the three corners of a triangular prism.   

The most important factor determining the qualitative properties of the population collapse is the Hill parameter $k$. A high value of $k$ causes an abrupt failure, while a low value of $k$ causes a gradual loss of cells.
The exact value of $k$ should depend on the microscopic mechanism by which the cooperative factors react with the damaging agent(s). Typically, large molecules are often secreted in low numbers but have large impacts, while small molecules individually have a lesser impact, but can be synthesized quickly. If, for example, a large molecule is secreted which helps reduce radicals or peroxides, which have been shown to damage cells and their DNA\cite{driessens}, a single cooperative molecule may react with dozens of hazardous agents. On the other hand, if the cooperative factor is a small molecule that chemically binds to the cell, or a molecule within, which causes genetic expression or the activation of different biochemical pathways, a large number of molecules would likely be needed to increase the cell's chance of survival. Therefore one should expect $k$ and $\phi_0$ to not necessarily be independent of each other.

Our experiments involved synthetic mammalian tissues with only one cell type, and motivated the assumptions that formed the basis of our theoretical analysis. However our conclusions may also hold true for the aging dynamics of a larger class of cooperative multicellular systems. Such systems may potentially include tumuors, biofilms, colonies and microbial consortia. Thus, we reiterate that aging is not a property of the individual components of a system, but an emergent characteristic of a strongly interacting, interdependent ensemble of components \cite{dcv}. However we should expect important qualitative differences in systems where a limiting cooperative factor is produced centrally and / or transported by a non-diffusive mechanism.

We should caution that there may be multiple cooperative factors with different diffusion lengths and different influences. Furthermore the survival of cells may depend on non-trivial combinations of cooperative factors. 

We expect multiple cooperative factors with similar diffusion lengths and influence not to change the mathematical structure of our theory, since they can be labeled as one, and summed into $\phi$. If the cells vitally depend on multiple cooperative factors, one can still get away with including the limiting one in $\phi$. However if there is a non-trivial dependence on substitutable combinations, e.g. if survival depends on (1 AND 2) or (3 AND 4), then eqn (\ref{hill}) should be modified appropriately, as
\begin{align}
P_i=\alpha_1\theta(\phi_1)\theta(\phi_2)+\alpha_1\theta(\phi_3)\theta(\phi_4)
\end{align}
where the $\theta$'s are Hill functions and concentrations of the four diffusive factors $\phi_1, \phi_2, \phi_3, \phi_4$ should be solved from eqn(\ref{dif}), with different $D$, $\gamma$, and $A$ constants, but will otherwise have the same functional form (\ref{totalcf}) each. We have not studied the implications of such complications, since we have not seen any experimental evidence for them. However our approach can be adapted accordingly, if such a complex arrangement of cooperative factors are observed to be a requirement for cell survival.

\section{Appendix: Methods}

\subsection{Isolation and Maintenance of Cardiac Fibroblasts} 
Neonatal rat cardiac-fibroblasts (CFs) were isolated from 2-day-old Sprague-Dawley rats (Charles River Laboratories). The rats were sacrificed by decapitation after CO2 treatment, and the hearts were immediately excised following the Institutional Animal Care and Use Committee (IACUC) guidelines of the University of Notre Dame. The excised hearts were digested in trypsin (Life Technologies) at 4C for 16h with gentle agitation. Next, the extracellular matrix of the hearts were further digested using collagenase type II (Worthington-Biochem) (37C) with several cycles of agitation and subsequent trituration. Then the mixture was filtered to separate the undigested tissue pieces and the filtrate which contains cardiomyocytes and CFs was seeded into a flask. Using the differential attachment of the two cell types, CFs were separated from cardiomyocytes at the end of 2h incubation. These CFs were passage 1 (P1) and they were maintained by media changes every 3 days. CF cultures were passaged at approximately 80\% confluency using trypsin-EDTA (0.05\%) (Life Technologies) and maintained in standard culture media (DMEM (Dulbeccoas modified eagle's media) supplemented with 10\% fetal bovine serum (Hyclone) and 1\% penicillin/streptomycin (Corning)). 

  \subsection{Preparation of Cell-laden Hydrogels and Determining Cell Survival} 
Passage 4 CFs were trypsinized and 1x$10^5$ cells were mixed 1:1 with 20\% (w/v) PEG (Jenkem)- RGD (Bachem) conjugated PEG (PEG-RGD) (PEG:PEG/RGD, 17:3) which contained 0.05\% (w/v in Phosphate Buffered Saline) final volume of photoinitiator (Irgacure 2959, BASF). Then, 10 microliters of the mixture was sandwiched between 125 micrometer thick spacers, and exposed to 6.9 mW/$ cm^2 $ UV irradiation for 20 sec. Triangle-shaped photomasks were used during UV exposure to control the shape of the hydrogels. These conditions allow us to model 3D tissues where there exists no diffusion limitations for nutrients. The hydrogels were exposed to stress by changing their media to standard culture media supplemented with 0.2 mM H2O2. Cell survival was determined using Live/Dead assay (Life Technologies). The hydrogels were stained using ethedium-homodimer-1 (stain for dead cells) every 24 hours and imaged using an inverted epifluorescence microscope (Zeiss, Hamamatsu ORCA flash 4.0). For experiments intended to investigate the propagation of the decay front the samples were stained at 0, 8, 16, 24, and 48 hours with both calcein-AM (stain for live cells) and ethidium homodimer-1 and imaged using an inverted epifluorescence microscope (Zeiss, Hamamatsu ORCA flash 4.0). Number of dead cells was determined using imageJ software and the decay penetration was determined by constructing a density heat map of contrast adjusted images in Mathematica where depth was measured as the straight-line distance from tip to the collapse of the decay wave front. As the propagation front has a non-linear dependence on curvature as a function of time, the straight line distance was used to create a measurement that could be compared across samples and conditions. We performed all cell culture experiments in triplicates. 

Data points represent the mean value, and error bars represent the standard deviation.\\\\
{\bf Ethics Statement:} All animal experiments were performed using protocols approved by Institutional Animal Care and Use Committee (IACUC) of University of Notre Dame, in accordance to the guidelines of National Institutes of Health, Office of Laboratory Animal Welfare.\\\\
{\bf Acknwoledgement:} This study is supported by National Science Foundation (NSF) Award PHY-1607643 and NSF-CAREER Award 1651385

\end{document}